\documentclass[12pt,preprint]{aastex}
\newcommand{\nsig}[1]{\mbox{{#1}$\sigma$}}

\shorttitle{A Photo-z Catalog from the RCS}
\shortauthors{Hsieh et al.}

\begin{document}

\title{A Photometric Redshift Galaxy Catalog from the Red-Sequence Cluster Survey}

\author{B. C. Hsieh\altaffilmark{1,2,6},
H. K. C. Yee\altaffilmark{3,6},
H. Lin\altaffilmark{4,6},
M. D. Gladders\altaffilmark{5,6}
}

\altaffiltext{1}{Institute of Astronomy, National Central University, 
No. 300, Jhongda Rd, Jhongli City, Taoyuan County 320, Taiwan, R.O.C.}
\altaffiltext{2}{Institute of Astrophysics \& Astronomy, Academia Sinica,
P.O. Box 23-141, Taipei 106, Taiwan, R.O.C. Email: bchsieh@asiaa.sinica.edu.tw}

\altaffiltext{3}{Department of Astronomy \& Astrophysics, University of 
Toronto, Toronto, Ontario, M5S 3H8, Canada. Email: hyee@astro.utoronto.ca}

\altaffiltext{4}{Fermi National Accelerator Laboratory, P.O. Box 500, 
Batavia, IL 60510. Email: hlin@fnal.gov}

\altaffiltext{5}{Carnegie Observatories, Pasadena, CA 91101.
Email: gladders@ociw.edu}

\altaffiltext{6}{Visiting Astronomer, Canada-France-Hawaii Telescope,
which is operated by the National Research Council of Canada, Le Centre 
National de Recherche Scientifique, and the University of Hawaii.} 

\begin{abstract}
The Red-Sequence Cluster Survey (RCS) provides a large and deep
photometric catalog of galaxies in the $z'$ and $R_c$ bands for ~90
square degrees of sky, and supplemental $V$ and $B$ data have been
obtained for 33.6 deg$^{2}$. We compile a photometric redshift catalog
from these 4-band data by utilizing the empirical quadratic polynomial
photometric redshift fitting technique in combination with CNOC2 and
GOODS/HDF-N redshift data. The training set includes 4924 spectral
redshifts. The resulting catalog contains more than one million galaxies with
photometric redshifts $< 1.5$ and $R_c < 24$, giving an rms scatter
$\sigma(\Delta{z}) < 0.06$ within the redshift range $0.2 < z < 0.5$
and $\sigma(\Delta{z}) < 0.11$ for galaxies at $0.0 < z < 1.5$.  We
describe the empirical quadratic polynomial photometric redshift
fitting technique which we use to determine the relation between
redshift and photometry. A kd-tree algorithm is used to divide up our
sample to improve the accuracy of our catalog. We also present a
method for estimating the photometric redshift error for individual
galaxies. We show that the redshift distribution of our sample is in
excellent agreement with smaller and much deeper photometric and
spectroscopic redshift surveys.

\end{abstract}


\section{Introduction}
The study of the formation and evolution of large-scale structure is
essential to our understanding of the universe. The typical method to
measure the distances of faint galaxies is to obtain their
spectroscopic redshifts. The spectroscopic method, however, is very
time consuming for surveying a large area with a deep limiting
magnitude, even when using a multi-object spectrograph. The less
accurate photometric redshift technique has been proved to be an
efficient and effective way to measure approximately the redshifts of
galaxies and to study their statistical properties and evolution
\citep[e.g.,][]{koo1985, connolly1995, GH1996, SLY1997, hogg1998,
WBT1998, FLY1999, benitez2000, csabai2000, budavari2000, rudnick2001,
csabai2003}. Basically, the technique is a transformation between a
set of observable parameters (e.g., magnitudes, colors) and estimates
of the physical properties of galaxies (e.g., redshift,
luminosity). Often the photometric redshift technique is applied to
multi-color deep imaging surveys over a small area (e.g., the Hubble
Deep Field, Sawicki et al.~1997 and Fern\'{a}ndez-Soto et al.~1999;
the Canada-France Deep Fields, \citealt{brodwin2003}; the COMBO-17
Survey, \citealt{wolf2003}; the Las Campanas Infrared Survey,
\citealt{chen2003}; the GOODS-S, \citealt{mobasher2004}). These
surveys have sufficient depth for measuring objects at high redshift
($z \gtrsim 1$), but they only probe small cosmological volumes which
might not provide a representative sample of the universe due to the
limited survey area. With more and more large multi-color imaging
surveys, larger and statistically more complete studies of galaxies
can be done. Although the application of photometric redshift
techniques to the Sloan Digital Sky Survey (SDSS) covers a large sky
area and provides a huge redshift sample, the photometric data are still
too shallow ($r$ $<$ 21) for galaxies at $z > 0.5$
\citep{csabai2003}. The Oxford-Dartmouth Thirty Degree Survey (ODTS,
\citealt{macdonald2004}) has both sufficient depth and a relatively
large area ($\sim$30 deg$^{2}$), but it does not have a large
spectroscopic training set to calibrate the photometric redshifts. A
deep multicolor survey with large area and proper spectroscopic data
will vastly improve on the current results.

The Red-Sequence Cluster Survey \citep[RCS,][]{GY2004} is a 4m-class imaging
survey in the $R_c$ and $z'$ bands covering $\sim 90$ deg$^{2}$ 
total in the northern and southern sky.
We have obtained follow-up observations
in $V$ and $B$ for a sub-area of the RCS. The four color photometry
($z'$, $R_c$, $V$, and $B$) allows us to cover a relatively large sub-area 
(33.6 deg$^{2}$) which samples a large cosmological volume
for galaxies up to redshift $\sim$ 1.5. 
This sub-area with four-color photometry overlaps with
the Canadian Network for Observational Cosmology (CNOC2) Field Galaxy Redshift 
Survey (Yee et al.~2000;~2005, in preparation) which provides several 
thousand redshifts which
can be used for the empirical photometric redshift technique
and provide verification of the photometric redshifts obtained.
By applying an empirical
quadratic polynomial photometric redshift fitting method to the RCS
photometric catalogs, a photometric redshift catalog containing about
one million galaxies at $z < 1.5$ with relatively small redshift errors
can be generated. 

In this paper we describe details of the RCS follow-up $V$ and $B$ 
observations and the techniques used to generate the photometric redshift 
catalog for the RCS survey. In \S\ref{data} we describe the data we 
used for the photometric redshift fitting and the photometric
calibrations for the $V$ and $B$ data. In \S\ref{trainingset} we provide a 
description of the spectroscopic training sets used in the photometric 
redshift fits. Section \ref{photoz-method} presents the photometric 
redshift technique we used. We describe the method used to estimate
the photometric redshift error in \S\ref{photoz-err}. We then 
demonstrate the results and present an example of the catalog in 
\S\ref{result}. In \S\ref{conclusion} we summarize the advantage 
of the photometric redshift catalog from the RCS and discuss possible 
follow-up studies based on the catalog.

\section{The Data}\label{data}
\subsection{Observations}\label{observations}
The RCS \citep{GY2004} was
designed to find clusters in a large volume, at high
redshift, and to low masses. The primary goals are to define a large 
sample of galaxy clusters at $z \sim 1$, hence providing a 
measurement of $\Omega_M$ and $\sigma_8$, and to perform studies of the 
evolution of galaxy clusters using a large complete sample.
Two filters ($z'$ and $R_c$) were chosen to find clusters at 
$0.4 < z < 1.4$ by applying the cluster red-sequence method 
\citep{GY2000}. There are 19 widely separated patches in both the northern 
and southern sky, covering $\sim 90$ deg$^{2}$ in total. The northern
half of the survey was observed with the CFH12K CCD camera on the 3.6m
Canada-France-Hawaii Telescope (CFHT), and the southern half was
obtained using the Mosaic II camera on the Cerro Tololo Inter-American
Observatory (CTIO) 4m Blanco telescope. For the CFHT RCS runs, each patch
consists of fifteen CFH12K pointings arranged in a slightly overlapping
grid of 3 $\times$ 5 pointings. The CFH12K camera is a 12k $\times$ 8k
pixel$^2$ CCD mosaic camera, consisting of twelve 2k $\times$ 4k pixel$^2$
CCDs (from MIT Lincoln Laboratories) arranged in a 6 $\times$ 2 grid.
It covers a 42 $\times$ 28 arcminute$^{2}$ area for the whole mosaic
at prime focus (f/4.18), corresponding to 0.2059 arcseconds per pixel.
We note that the average East-West (X) gap between the CCDs is 7.8 arcseconds,
and the average North-South (Y) gap between the CCDs is 6.8 arcseconds. 
The observation dates are from May 1999 to January 2001. 
The typical seeing is 0.62 arcsec for $z'$ and 0.70 arcsec for $R_c$.
The integration times are typically 1200s for $z'$ and 900s 
for $R_c$, with average \nsig{5} limiting magnitudes of 
$z'_{AB} = 23.9$ and $R_c = 25.0$ (Vega) for point sources,
adoping an aperture of diameter 2.7 arcsec (on average).
The data and the photometric 
techniques are described in detail in \citet{GY2004}.

\begin{figure}[t]
\includegraphics[angle=270,width=0.9\textwidth]{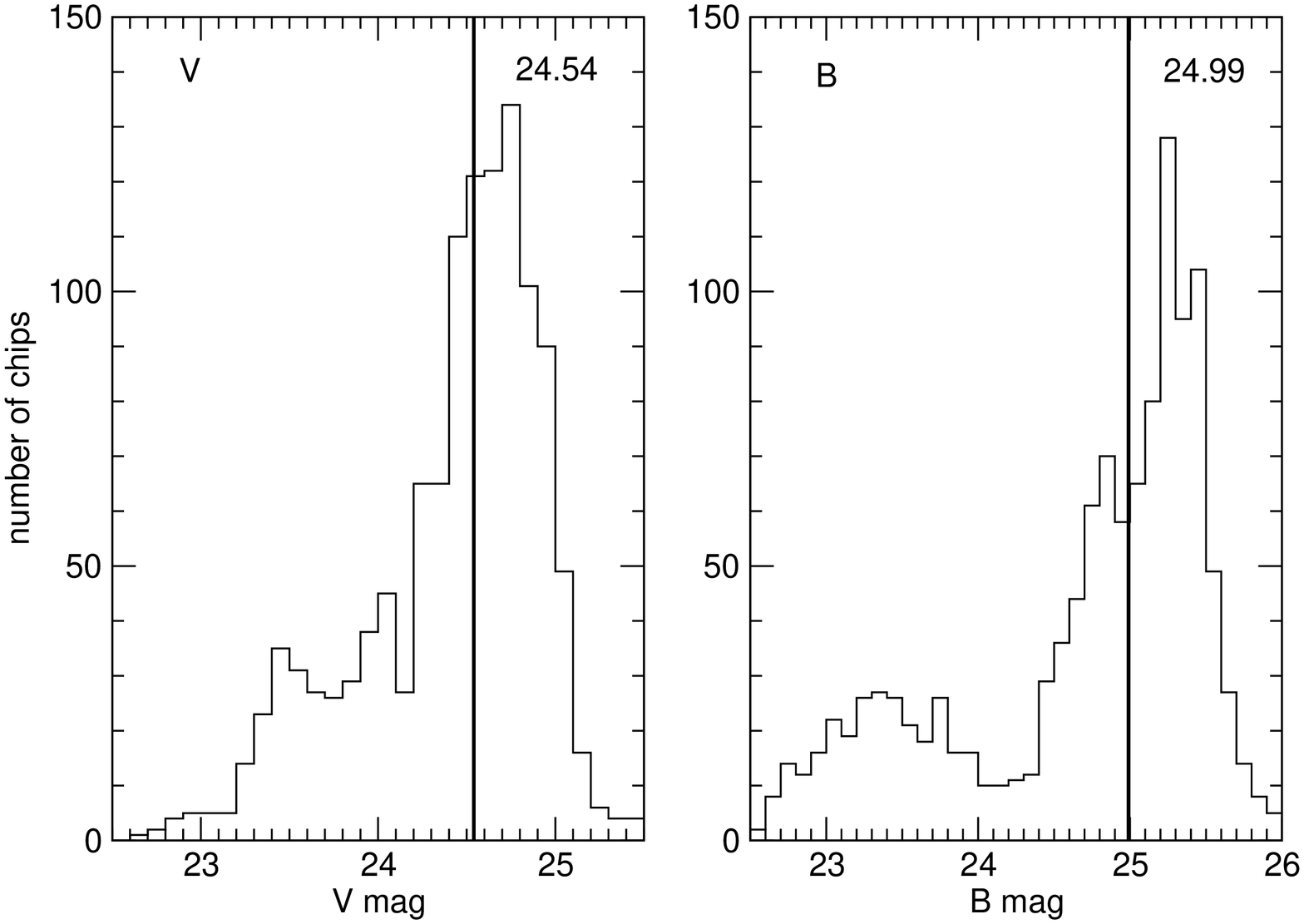}
\caption{Histograms of $V$ and $B$ \nsig{5} limiting magnitudes. 
The distributions of \nsig{5} point source magnitudes from 
individual chips for the $V$ filter and $B$ filter are shown in the left panel
and the right panel, respectively. The thick lines indicate the median value of 
the limiting magnitudes for $V$ and $B$, which are 24.54 and 24.99, 
respectively. Note that a number of pointings have very shallow $B$ limits
due to observations done in poor seeing.
\label{vb_lim} }
\end{figure}

The follow-up observations in $V$ and $B$ were obtained using the CFH12K 
camera and cover 33.6 deg$^{2}$ ($\sim 75$\% complete 
for the original CFHT RCS fields).
The follow-up $V$ and $B$ observations contain 108 pointings spread
over the ten CFHT RCS patches. The pointings of the CFHT RCS patches are 
shown in Table~\ref{patch}. For each pointing, a single exposure in each field 
was taken without dithering. The observation dates are from May 2001 to 
June 2002. Most data were taken in three runs in queue mode: 
01AC44, 01BC25, and 02AC25. The data for another 29 pointings were 
taken using the observing time (run 01BC28) shared with the EXPLORE project
\citep{yee2002}. Some data were also taken during two CNOC2 photometric
redshift runs using CFH12K in January 2000 and December 2000. We had 16 
photometric nights out of 50 nights in which data were taken.
Because some data were taken under non-photometric conditions, short
exposures were obtained in photometric weather for calibration.
The typical seeing is 0.65 arcsec for $V$ and 0.95 arcsec for $B$.
The average exposure times for $V$ and $B$ are 480s and 840s, respectively,
and the median \nsig{5} limiting magnitudes (Vega) for point sources are 24.5 
and 25.0, respectively, with an adopted aperture of diameter 2.9 arcsec (on average).
We present histograms of the $V$ and $B$ 5$\sigma$ limiting 
magnitudes in Figure~\ref{vb_lim}.

\clearpage

\begin{deluxetable}{ccccccc}
\tabletypesize{\scriptsize}
\rotate 
\tablecolumns{7}
\tablewidth{0pt}
\tablecaption{CFHT RCS Patches\label{patch}}
\tablehead{
\colhead{Patch number} &
\colhead{Patch name} & \colhead{RA(2000)} & \colhead{DEC(2000)} &
\colhead{Area (deg$^{2}$) with $R_c$ and $z'$} & 
\colhead{Area with four filters} & \colhead{Notes} }
\startdata
01 & 0226+00 & 02 26 07.0 & +00 40 35 & 4.81 & 3.99 & CNOC2 Patch${^a}$\\
02 & 0351-09 & 03 51 20.7 & -09 57 41 & 4.79 & 4.30 &             \\
03 & 0926+37 & 09 26 09.6 & +37 10 12 & 5.59 & 4.85 & CNOC2 Patch${^a}$\\ 
04 & 1122+25 & 11 22 22.5 & +25 05 55 & 4.78 & 4.72 &             \\
05 & 1327+29 & 13 27 41.9 & +29 43 55 & 4.54 & 1.34 &  PDCS Patch${^b}$\\
06 & 1416+53 & 14 16 35.0 & +53 02 26 & 4.53 & 3.04 & Groth Strip \\
07 & 1449+09 & 14 49 26.7 & +09 00 27 & 4.17 & 2.01 & CNOC2 Patch${^a}$\\
08 & 1616+30 & 16 16 35.5 & +30 21 02 & 4.26 & 4.16 &             \\
09 & 2153-05 & 21 53 10.8 & -05 41 11 & 3.43 & 2.96 & CNOC2 Patch${^a}$\\
10 & 2318-00 & 23 18 10.7 & -00 04 55 & 4.84 & 2.23 &             \\
\enddata
\tablerefs{${^a}$\citet{yee2000}, ${^b}$\citet{postman1996}}
\end{deluxetable}

\clearpage

\subsection{Photometric Data Reduction}\label{data-reduction}
Photometric data reduction including object finding, star-galaxy
classification, and aperture photometry was performed using the 
program PPP, described in detail in \citet{yee1991} and \citet{yee1996}.

The RCS $z'$ and $R_c$ object finding and photometry were already done
as part of the primary cluster finding program,
and are described in detail in \citet{GY2004}. To simplify the photometric
measurement procedure, we used the pixel positions of objects in the 
$R_c$ image to do photometry for $V$ and $B$. To achieve this, the 
position of each object in $V$ and $B$ must be accurately determined 
to within two pixels in radius relative to the $R_c$ image, sufficiently
close for PPP to do a recentroiding of the object position before 
performing photometry.
However, the follow-up observations in $V$ and $B$ began a semester
after the observations in $z'$ and $R_c$ were finished. The
CFH12K CCD was taken off and re-installed on the prime focus for
each run, and the camera was not at the identical position after each
re-installation. Hence, the $V$ and $B$ images are rotated slightly
relative to the $z'$ and $R_c$ images for the same field. The telescope
pointing is also not identical. To deal with these problems, we 
match the $V$ and $B$ images to the $R_c$ image before performing photometry. 

Because we did not dither to make one large image per pointing, we  
match the images chip by chip. The transformation function between the 
$V$ and $B$ images and the $R_c$ images can be easily determined 
from bright reference objects. However, the new transformed coordinates 
for each pixel are almost always not integer numbers after applying the
transformation function with non-integer shifts and rotation.
Therefore re-sampling is needed. To avoid degradation of the image quality
and to preserve the Poisson characteristics of the photon noise
in the images, we use the nearest
neighbor re-sampling algorithm, which also preserves flux. We use the 
closest integer coordinate of the transformed non-integer coordinate as the 
new coordinate for each original pixel. However, two original pixels may
produce two transformed non-integer coordinates which may have the same 
nearest integer coordinate. In other words, 
one new pixel may be assigned fluxes from two original pixels because of 
round off in the pixel coordinates. In addition, some new coordinates
may have zero flux because they are not the nearest coordinates
for any original pixels. The photometry results would not be correct
when either condition occurs within a photometry aperture. 
We try to deal with the problem by subtracting the local background
from a pixel containing double fluxes because that pixel will also contain
double the background counts. On the other hand, we put the counts of the 
local background into those pixels with zero flux. These procedures
minimize the errors on the photometry. 
We note that on average, only approximately 500 pixels 
(ranging from 0 to 3,000 depending on the rotation angle) are affected for 
each chip. Given that the total number of objects in each chip is
$\sim 5,000$ and that the average diameter of the photometry aperture is 
14 pixels, we estimate that only $\sim 50$ pixels in photometry apertures 
are affected in each chip; i.e., only one object out of 100 is affected and the
effect has been minimized by the method we described above
(the magnitude difference before and after the transformation for an
affected object is less than 0.02 magnitude). 

\begin{figure}[t]
\includegraphics[angle=270,width=0.9\textwidth]{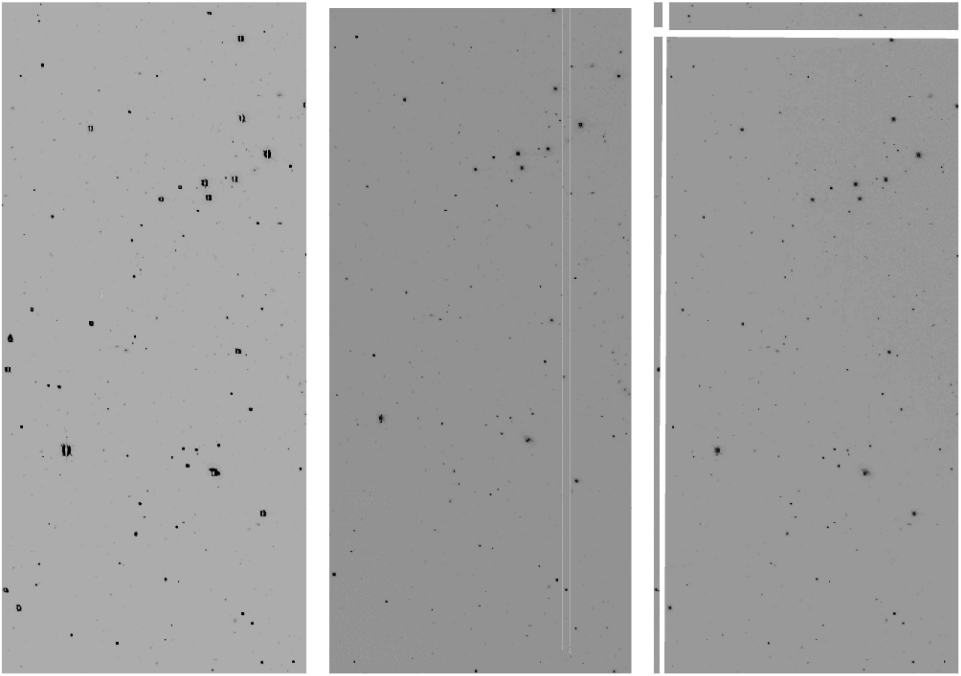}
\caption{ An example of resampled images from field 2153A1/CCD Chip07. 
The left image in $R_c$ is used as the reference image. The central
image is the original (before matching) image in $V$. The right
image is the shifted/rotated image in $V$ made by matching to the $R_c$
image. Note that the upper part and left part of the right image
are copied from the nearby chips to keep the completeness of the
$V$ image, and the gaps are not filled with any counts here for 
clarity of presentation. The bad columns in the original $V$ image are
fixed for the shifted/rotated $V$ image by using a linear interpolation
algorithm.\label{matching-image} }
\end{figure}

Additionally, the transformed 
image of each chip often overlaps with several nearby chips of the 
original image, and the matched regions 
from other chips are also cropped and included in the transformed image.
We note that within each pointing, images from all the chips have
been scaled to have the same zeropoint. An example of a re-sampled $V$ image is 
shown in Figure~\ref{matching-image}. The gaps between chips are also
filled with counts of the local background. Objects falling into the gaps
are lost. The final transformed $V$ and $B$ images are matched with the 
$R_c$ images for the same fields. The typical shift is between a few pixels 
to one hundred pixels. The typical rotation angle is about 0.03 degree and
the largest angle is less than one degree. 

After the matching is done, we 
apply the coordinate file of objects from the $R_c$ image to perform the 
aperture photometry procedure. Since the same coordinate file is used for 
the data in different filters, the object ID of each object in different
filters of the same pointing remains the same. The multi-color
photometric catalog can be generated very easily by just combining
the photometric results object by object directly. The colors are measured
using identical apertures for all four filters, as prescribed in
\citet{yee1991}.

\begin{figure}[t]
\includegraphics[angle=270,width=0.9\textwidth]{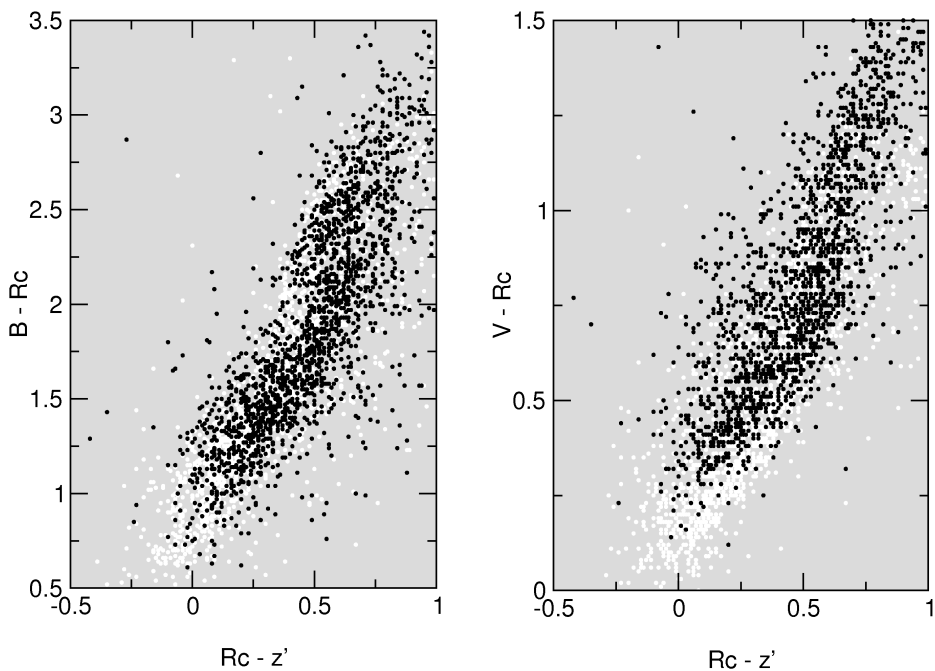}
\caption{Color-color diagrams of 0226A3 and 0226C1.
The left panel is the $B - R_c$ vs. $R_c - z'$ diagram. The right panel is
the $V - R_c$ vs. $R_c - z'$ diagram.
The white dots and black dots indicate extended sources with $R_c < 21.5$
in 0226A3 and 0226C1, respectively. The distributions of galaxies in the 
two pointings in the color-color diagrams do not match, especially for
$V - R_c$ vs. $R_c - z'$. It is evident that the original photometric 
calibrations are not sufficiently accurate, and some additional recalibrations 
have to be performed.
\label{color-color} }
\end{figure}

\begin{figure}[t]
\includegraphics[angle=270,width=0.9\textwidth]{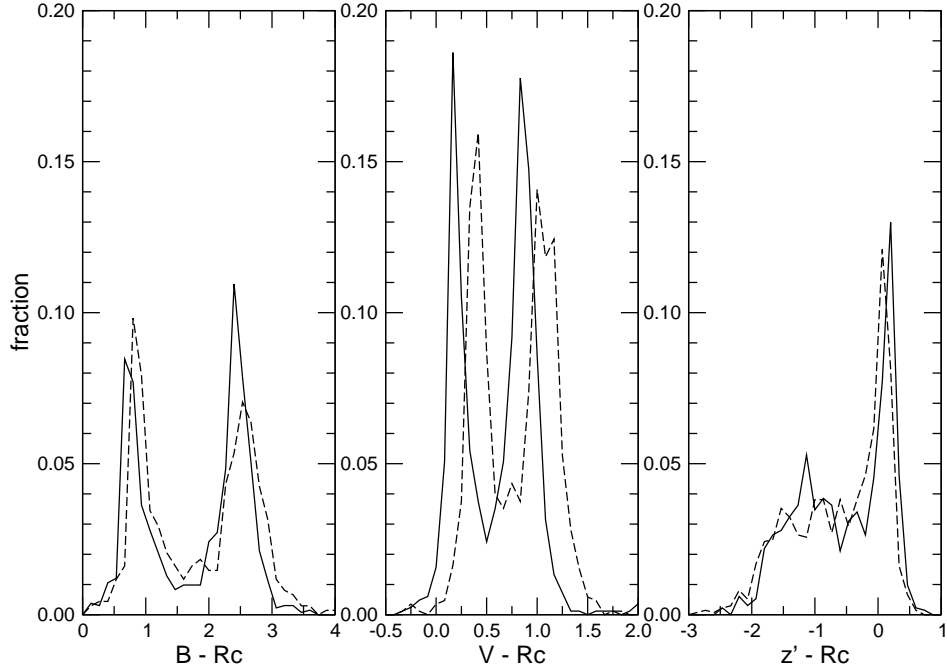}
\caption{Color histograms of 0226A3 and 0226C1.
The panels from left to right are for the colors $B - R_c$, $V - R_c$, and
$z' - R_c$, respectively. The solid line indicates the data from 0226A3 and
the dashed line indicates the data from 0226C1. Only stars with
$18 < R_c < 22$ are used. All histograms are normalized. In the
three panels the color histograms for the two pointings are similar in
shape but are offset from each other. By assuming that the 
photometric calibrations for $R_c$ are all correct and by
choosing the pointing 0226A3 as the reference pointing, the magnitudes 
of $B$ and $V$ should be brighter and the $z'$ magnitude should be a little
fainter for the 0226C1 pointing. Shifts of $-0.11$, $-0.18$, and $0.09$ mag
need to be applied to the $B$, $V$, and $z'$ magnitudes, respectively, 
of the 0226C1 pointing in order to make the color histograms match.
\label{color-histo} }
\end{figure}

The preliminary $V$ and $B$ photometric calibrations (zero-point, airmass, 
and color terms) were taken directly from the CFHT Elixir solutions in the 
image FITS file headers. The Elixir
\footnote{\url{http://www.cfht.hawaii.edu/Instruments/Elixir/home.html}}
system at CFHT provides real-time data quality
assessment, end-of-run detailed calibration analysis, and image 
pre-processing and meta-data compilation for data distribution.
Data taken under bad weather conditions are calibrated by short exposures 
taken in photometric nights. However, we find that the color 
($V - R_c$ or $B - R_c$) vs. color ($R_c - z'$) diagrams have offsets from 
pointing to pointing, and sometimes the offsets are as large as $\sim 0.3$ 
magnitude, which implies that the photometric calibrations are not sufficiently 
accurate. An example of the differences between two pointings on 
the color-color diagram is shown in Figure~\ref{color-color}.
Hence, we use the following method to recalibrate the photometry.
According to a recent star count study \citep{parker2003}, although
the densities of stars would be different for different fields, the 
normalized color distributions for bright stars are very similar 
from field to field as long as the selected fields are sufficiently large.
Based on this result, we recalibrated the $z'$, $V$, and $B$ photometry by
using the $z' - R_c$, $V - R_c$, and $B - R_c$ color distributions, 
assuming that the photometry of $R_c$ calibrated by standard stars
is correct \citep{GY2004}. Stars with $18 < R_c < 22$ are selected to 
determine the magnitude offsets on a pointing-to-pointing and patch-to-patch basis.
Figure~\ref{color-histo} represents an example of the difference between 
the color histograms of two pointings. 
For the pointing-to-pointing recalibration within each patch, the pointing with
the least scatter in the color-color diagrams ($V - R_c$ or $B - R_c$ vs.
$R_c - z'$) is chosen to be the reference pointing. (The reference pointings
are 0226A3, 0351C1, 0926B4, 1122C5, 1327B2, 1416C1, 1449B2, 1616C5, 2153A1,
and 2318B1 for the 10 patches.) The magnitude offsets in $z'$, $V$, and $B$ 
are computed using cross-correlation of the distributions of 
$z' - R_c$, $V - R_c$, and $B - R_c$, respectively, between the reference 
pointing and the test pointing. The patch-to-patch photometry
recalibrations are performed after the pointing-to-pointing recalibrations
are done. Patch 0926 is chosen to be the reference patch. The same
recalibration procedure as the pointing-to-pointing recalibration is 
applied to the patch-to-patch recalibration, and all the pointings from
each patch are used during the patch-to-patch recalibration procedure.

\begin{figure}[t]
\includegraphics[angle=270,width=1.0\textwidth]{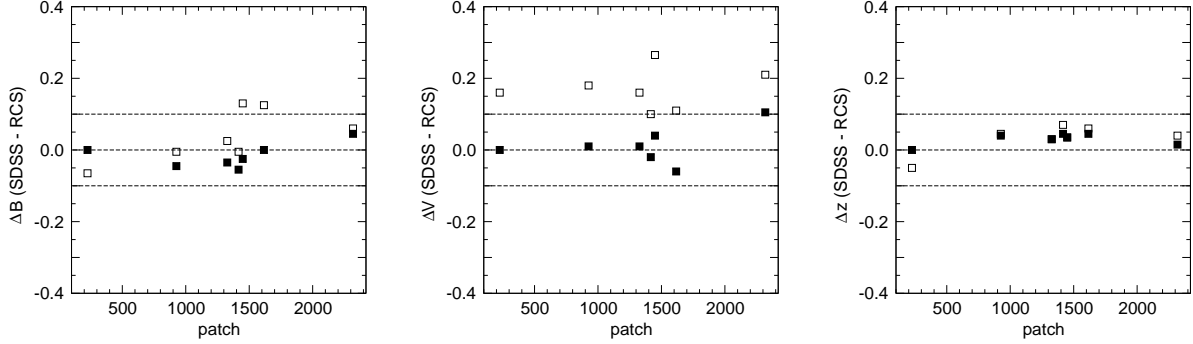}
\caption{The zeropoint differences $\Delta$mag between RCS and SDSS 
in $B$, $V$, and $z'$ (left to right) for overlapping patches;
see text for details.
We arbitrarily set $\Delta$mag $= 0$ for patch 0226 
{\it after} recalibration.
Note that we do not calibrate the RCS photometry to the SDSS
photometry. The open boxes and filled boxes indicate the
Elixir-calibrated results and the recalibrated results, respectively. The 
recalibrated results show less scatter than the Elixir-calibrated ones.
\label{sdss-offset} }
\end{figure}

\begin{figure}[t]
\includegraphics[angle=270,width=1.0\textwidth]{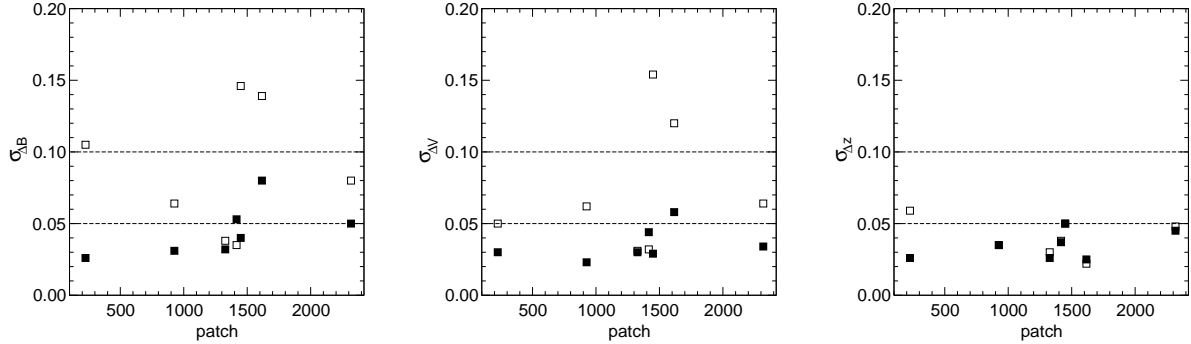}
\caption{The standard deviations of the zeropoint difference
$\Delta$mag between RCS and SDSS for the set of pointings within 
each RCS patch; see text for details.
$B$, $V$, and $z'$ are shown (left to right).
The open boxes and filled boxes indicate the
Elixir-calibrated results and the recalibrated results, respectively. The
inter-pointing recalibrations for each patch reduce the standard deviations
for $V$ and $B$ significantly.
\label{sdss-sd} }
\end{figure}

We compare the recalibrated photometry of galaxies with the overlapping
published
SDSS\footnote{Funding for the Sloan Digital Sky Survey (SDSS) has been
provided by the Alfred P. Sloan Foundation, the Participating Institutions, the National Aeronautics and Space Administration, the National Science
Foundation, the U.S. Department of Energy, the Japanese Monbukagakusho,
and the Max Planck Society. The SDSS Web site is \url{http://www.sdss.org/.}
The SDSS is managed by the Astrophysical Research Consortium (ARC) for the
Participating Institutions. The Participating Institutions are The University
of Chicago, Fermilab, the Institute for Advanced Study, the Japan Participation
Group, The Johns Hopkins University, the Korean Scientist Group, Los Alamos
National Laboratory, the Max-Planck-Institute for Astronomy (MPIA), the
Max-Planck-Institute for Astrophysics (MPA), New Mexico State University,
University of Pittsburgh, Princeton University, the United States Naval
Observatory, and the University of Washington.}
database as a check on our recalibration procedure.
Patches 0226, 2318, 0926, 1416, 1449, and 1616 overlap
with the SDSS Data Release 3 \citep[DR3;][\url{http://www.sdss.org/dr3/}]{dr3}.
We match the
objects between RCS and SDSS, and use the following equations to determine
the relation between the two different photometry systems:
\begin{eqnarray}\label{comparison}
B = g + slope_B(g-r) + \Delta{B} \nonumber \\
V = g + slope_V(g-r) + \Delta{V} \nonumber \\
R_c = r + slope_{R_c}(r-i) + \Delta{R_c} \nonumber \\
z' = z + slope_{z'}(r-i) + \Delta{z'},
\end{eqnarray}
where the $slope$ term is the coefficient of the color term in the
transformation,
and $\Delta$mag, where mag $= B, V, R_c, z'$, is the magnitude zeropoint
difference between the RCS and SDSS photometry systems.
We compare RCS magnitudes against SDSS model magnitudes.
The quality of the recalibration can be verified by checking
the consistency of the $\Delta$mag values between different RCS patches. In
Figure~\ref{sdss-offset} we plot $\Delta$mag for $B$, $V$, and $z'$
for those RCS patches overlapping the SDSS, where we have arbitrarily set
$\Delta$mag $=0$ for patch 0226 {\it after} recalibration.
For these patch-to-patch comparisons we have ignored
the slope terms above and simply set $\Delta$mag to the median offset
between the respective RCS and SDSS magnitudes for each patch.
The open boxes indicate $\Delta$mag using Elixir calibrations,
and the filled boxes indicate $\Delta$mag after recalibration. The
recalibrated data have less scatter between patches than the Elixir-calibrated
data in all the panels, especially for $B$. Figure~\ref{sdss-sd} shows the
standard deviations of the $\Delta$mag values for $B$, $V$, and $z'$
among the different pointings within each RCS patch.
Here we first fit the above transformation equations to each patch
as a whole, transform the SDSS magnitudes accordingly, and
then compute the median offset between RCS magnitudes and transformed SDSS
magnitudes for each pointing within a patch.
The standard deviations of these pointing-by-pointing offsets
are plotted in Figure~\ref{sdss-sd}.
The open boxes and the filled boxes again indicate the Elixir-calibrated
results and the recalibrated results, respectively.
The pointing-to-pointing recalibrations
reduce the standard deviations dramatically for $V$ and $B$. This RCS-SDSS
comparison proves that recalibration using the color distributions of
bright stars significantly improves the calibrations of the $V$ and $B$
photometry over the original Elixir-based calibrations,
and that most of the original photometric calibrations (as performed by RCS,
\citealt{GY2004}) for $z'$ are already good.

Galactic extinction has to be applied to the data before doing
photometric redshift fitting. The extinction values are calculated
according to prescriptions from \citet{cardelli1989}, \citet{odonnell1994},
and \citet{schlegel1998}. The galactic extinction values for the four filters 
for each patch are shown in Table~\ref{extinction}. 
Note that our procedure will tend to result in slightly wrong
galactic extinction corrections for the $B$, $V$, and $z'$ filters, 
since the stars used in the recalibration procedure already suffer 
extinction to some extent and the recalibration procedure will
have corrected for some of the extinction differences among the patches.
However, the error we make is reduced because we use colors in
the recalibration and because we use relative shifts among the patches.
For example, if we assume that the stars we use already suffer the full 
amount of galactic extinction, then for patch 2318, which
has the largest $E(B-V)$ value, we would have made only errors of 
-0.049, -0.022, and 0.024 mag for $B$, $V$, and $z'$, respectively.  
Moreover, the comparison against SDSS data 
described above and shown in Figure~\ref{sdss-offset} indicates
that the extinction of stars does not cause large errors in our 
recalibration procedure.

\clearpage

\begin{deluxetable}{crrrrrrrrrr}
\tablecolumns{11}
\tablewidth{0pt}
\tablecaption{Galactic Extinction\label{extinction}}
\tablehead{
\colhead{patch name} & \colhead{0226} & \colhead{0351} & \colhead{0926} &
\colhead{1122} & \colhead{1327} & \colhead{1416} & \colhead{1449} &
\colhead{1616} & \colhead{2153} & \colhead{2318} }
\startdata
E(B-V) & 0.036 & 0.043 & 0.012 & 0.018 & 0.012 & 0.010 & 0.029 &
0.038 & 0.035 & 0.044 \\ 
$A_B$ & 0.140 & 0.165 & 0.048 & 0.068 & 0.046 & 0.037 & 0.112 &
0.146 & 0.135 & 0.167 \\
$A_V$ & 0.108 & 0.127 & 0.037 & 0.053 & 0.035 & 0.028 & 0.086 &
0.113 & 0.105 & 0.129 \\ 
$A_{R_c}$ & 0.083 & 0.098 & 0.029 & 0.040 & 0.027 & 0.022 & 0.066 &
0.087 & 0.080 & 0.099 \\
$A_{z'}$ & 0.053 & 0.063 & 0.018 & 0.026 & 0.018 & 0.014 & 0.043 &
0.056 & 0.052 & 0.064 \\
\enddata
\end{deluxetable}

\clearpage

\section{Spectroscopic Training Sets}\label{trainingset}
Since we use an empirical quadratic polynomial fit to
estimate photometric redshifts for the RCS data, we
need spectroscopic data which overlap with the RCS fields to
create a training set. We primarily use the spectroscopic data from the
CNOC2 project, but we also include GOODS/HDF-N data to improve the
limiting magnitude and redshift range of the training set. All the
data included in our training set are described in detail below.

\subsection{CNOC2} The Canadian Network for Observational
Cosmology (CNOC2) Field Galaxy Redshift Survey 
(Yee et al.~2000;~2005, in preparation) is
a spectroscopic/photometric survey of faint galaxies. There are
four widely separated patches named 0226, 0926, 1449, and 2153.
The data were obtained using the Multi-Object Spectrograph on the CFHT.
The survey covers over 1.5 deg$^{2}$ of sky with a total sample
of $\sim 6200$ redshifts with $R_c \lesssim 22$. The CNOC2 survey 
used a band-limiting filter for the spectroscopic observations to increase the 
number of objects observed. This produces an effective redshift range for the 
statistically complete sample of 0.12-0.55, and 0.0-0.68 for emission-line 
galaxies. The nominal statistical completeness magnitude is $R_c = 21.5$, but 
there are objects as faint as $R_c \sim 22.5$ in the redshift sample.

Fifteen RCS pointings, distributed over four RCS patches and including 
four-filter photometric data, overlap the CNOC2 fields. We match the
objects from the two surveys and create a training set containing the
photometric data from the RCS and the spectroscopic redshifts from
CNOC2. There are 3,130 objects in this training set.

\subsection{GOODS/HDF-N} Compared to the spectroscopic limit of
CNOC2, the photometric data of RCS are much deeper (100\% completeness
limit $R_c \simeq 24.2$ for RCS compared to the spectroscopic completeness
limit $R_c = 21.5$ for CNOC2). Furthermore, the CNOC2 spectroscopic sample
has a nominal redshift limit of $\sim$ 0.55 due to the use of a
band-limiting filter, and so does not provide
a good training set for objects with $R_c \lesssim 21.5$ at $z \gtrsim 0.6$.
A deeper dataset (limiting magnitude of $R_c \sim 24.0$) with higher
redshifts ($z > 0.6$) is thus needed in addition to the original training set
to improve redshift estimates for high-$z$ objects in the RCS.
Thus, we choose the GOODS/HDF-N data to be the additional training
set for high-$z$ and fainter objects.

The Great Observatories Origins Deep Survey
\citep[GOODS;][]{giavalisco2004} is a survey based
on multi-band imaging data obtained with the Hubble Space
Telescope (HST) and the Advanced Camera for Surveys (ACS). It
covers two fields (HDF-N and CDF-S) with roughly 320 arcmin$^{2}$
total area.
For our high-redshift, faint-magnitude training set,
we use publicly available $BVRz'$ photometry and spectroscopic redshifts 
for the GOODS/HDF-N field.
The photometry comes from the ground-based Hawaii HDF-N data
set of \citet{capak2004}, which was obtained with the Subaru 8.3m telescope
and has \nsig{5} limiting magnitudes $B_{AB} = 26.9$, $V_{AB} = 26.8$,
$R_{AB} = 26.6$, and $z'_{AB} = 25.4$ measured in 3" diameter
apertures, and typical integration times of 600s, 1200s, 480s, and
180s/240s, respectively. The spectroscopic redshift data
for the GOODS/HDF-N field are from the samples of 
\citet{wirth2004} and \citet{cowie2004}, obtained using the 
Keck 10m telescope.

After combining and matching objects in the Hawaii HDF-N photometric
catalog to the GOODS/HDF-N spectroscopic catalog, a training
set containing 1,794 objects is generated. This additional training set
is not only deeper than the RCS+CNOC2 set, but it also contains many more
objects with higher redshifts ($0.5 < z < 1.5$). To combine the CNOC2 and
GOODS training sets, we also need to offset the magnitudes of the GOODS data to
match the zero-points of the RCS data. 
We did not use stars to derive the magnitude offsets because
the field is too small to contain a statistically sufficient number
of stars for the purpose. We first apply galactic extinction
corrections to the GOODS/HDF-N data 
($-0.018$, $-0.028$, $-0.036$, and $-0.047$ for $z'$, $R_c$, $V$, and $B$), 
and then compute the difference in the color distributions for 
galaxies with $R_c < 21.5$ between the RCS and GOODS/HDF-N samples.
By assuming $R_c$ is always correctly calibrated, the magnitude 
offsets for $z'$, $V$, and $B$ can be determined. 
Magnitude offsets $-0.1$, $-0.05$, and $0.0$ for $z'$, $V$, and $B$, 
respectively, 
are applied to the GOODS/HDF-N data. The final training set
created by combining the RCS+CNOC2 and the GOODS/HDF-N
data includes 4924 objects.  This combined training set will improve 
the accuracy of photometric redshifts, especially for objects at
fainter magnitudes and higher redshifts.

\section{Photometric Redshift Method}\label{photoz-method}
Photometric redshift techniques have been developed for decades 
\citep[e.g.,][]{koo1985},
but there are two primary approaches to estimating the photometric redshift. 
One way is to compare the photometric data against templates generated from 
models or from a real spectral energy distribution database
(e.g., Hogg et al.~1998; Fern\'{a}ndez-Soto et al.~1999).
The other way is to find the empirical 
relation between the photometric data and the redshift identified 
from the spectroscopic data, e.g., empirical polynomial fitting
(Connolly et al.~1995). The empirical method is especially effective
(e.g., see \citealt{csabai2003}) when there is a large spectroscopic redshift
data set available, as in the case of our RCS data set.

We use empirical quadratic polynomial fitting to estimate photometric 
redshifts for the RCS data. First we need to generate a subset called a 
``Training Set,'' which includes information on object ID, spectral redshift, 
and photometric data for each filter (see Section~\ref{trainingset}). 
We then fit this subset with the following second order equation using 
least-squares fitting:
\begin{eqnarray}\label{photozeqn}
redshift = const + {a_0}B^{2} + {a_1}V^{2} +
{a_2}{R_c}^{2} + {a_3}z'^{2} \nonumber \\
+ {b_0}B + {b_1}V + {b_2}{R_c} + {b_3}z' + {c_0}BV \nonumber \\
+ {c_1}B{R_c} + {c_2}Bz' + {c_3}V{R_c} + {c_4}Vz' + {c_5}{R_c}z'.
\end{eqnarray}
Including the constant term, 15 parameters are derived from the
fit. The above formula describes the empirical relation
between the photometric data and the spectroscopic redshift. 
By applying this formula to the photometric data of an object, 
an estimated photometric redshift for that object is readily obtained. 

\begin{figure}[t]
\includegraphics[angle=270,width=0.9\textwidth]{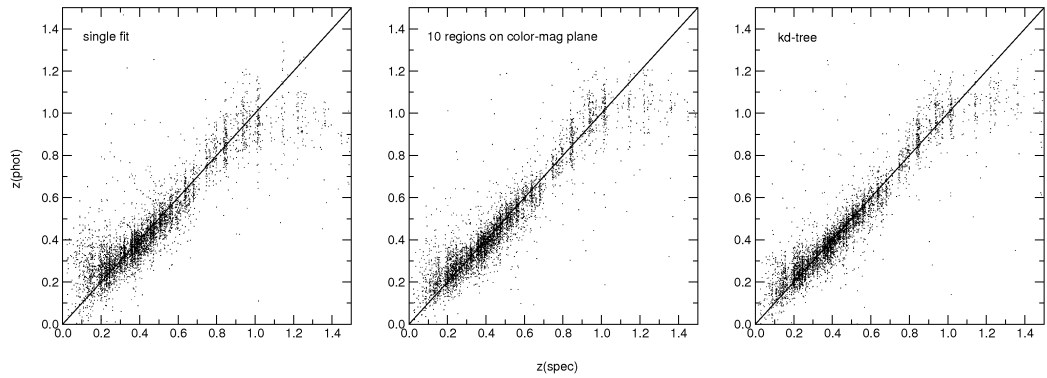}
\caption{Photometric redshift vs. spectroscopic redshift for the RCS/CNOC2
and GOODS/HDF-N data in our training set. The panels
from left to right use: (1) brute force single fit for 
all data, (2) cutting into 10 regions in the color-magnitude plane, 
and (3) the kd-tree method with 32 cells. The result using the
kd-tree method has the least systematics at both the high-$z$ and low-$z$ ends,
as well as the least scatter of the three methods.
\label{photoz_comp}}
\end{figure}

One can use a brute-force single fit for all the data, but fitting all galaxies 
with a single quadratic formula is not optimal, since different types of 
galaxies may have different fitting parameters in the quadratic formula. 
The single fit method gives a result with large scatter and systematic 
deviations (see left panel, Figure~\ref{photoz_comp}). To improve the 
photometric redshift results, different types of galaxies 
should be fit separately. We describe below two methods of separating
galaxies into different samples.

\begin{figure}[t]
\includegraphics[angle=270,width=0.9\textwidth]{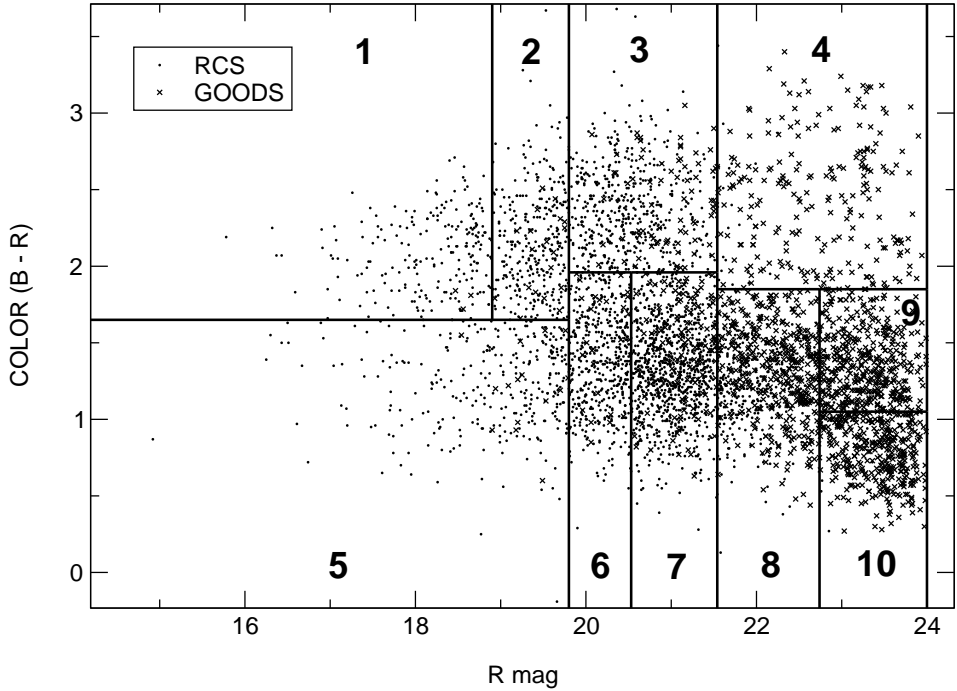}
\caption{Color ($B - R_c$) vs. magnitude ($R_c$) diagram showing the 
layout of the regions used for the data separation. 
The dots indicate the RCS+CNOC2 data 
and the crosses indicate the GOODS/HDF-N data in our training set.
The galaxies are roughly in two loci on the color-magnitude plane with
a gap around $B - R_c$=1.8 to 2.0. The early-type galaxies dominate the
redder (upper) locus while the late-type galaxies dominate the bluer
(lower) locus. We separate the upper locus into four regions
(1, 2, 3, and 4) and the lower locus into six regions (5, 6, 7, 8, 9, and 10)
for the photometric redshift fits. Note that this cutting method is not
used for the final catalog.
\label{color-mag}}
\end{figure}

Color is one of the important signatures for identifying the 
galaxy type. However, it is subject to K-corrections for galaxies at
different redshifts. Thus, other parameters have to be used to break the 
color-redshift degeneracy. Roughly speaking, more distant galaxies have 
fainter magnitudes, and for a reasonable range of galaxy types, they also
have redder observed colors. 
Hence by using some appropriate boundaries to divide 
galaxies in the color-magnitude plane, we can mimic very roughly the effect 
of separating galaxies of different types at different redshifts.
Figure~\ref{color-mag} presents a color ($B - R_c$) vs. magnitude ($R_c$)
diagram. The sample data consist of RCS+CNOC2 (dots) and GOODS 
(crosses) galaxies with $R_c < 24$ in our training set
(see \S\ref{trainingset} for a detailed description of the training set). 
There is a gap around $B - R_c$=1.8 to 2.0 which roughly separates
early-type galaxies (redder) and late-type galaxies (bluer). We divide
the color-magnitude plane into ten parts. Regions 1-4 are for 
redder galaxies, and region 5-10 are for bluer galaxies.
This method produces a result with smaller scatter and systematic deviations than 
the single fit method (middle panel, Figure~\ref{photoz_comp}). 
However, the $B - R_c$ color uses filters
that are too blue to make a good separation for galaxies 
with redshifts larger than 0.6, where the 4000\AA \ break will be shifted to
$>$ 6500\AA. To solve this problem, a redder filter ($z'$) has to be used 
in the separation criteria to produce a better result.

In our second method of separating the training set, we add one more color 
($R_c - z'$) to the original two-dimensional color-magnitude plane to form 
a three-dimensional color-color-magnitude space. The kd-tree algorithm, which
uses median values to divide up the data points in a k-dimensional space 
successively \citep{bentley1979}, is used to separate galaxies in our
three-dimensional space. We use a kd-tree depth of five, so that the
space is separated into 32 cells. Each cell contains about 150 objects.
The photometric redshift fitting procedure is applied to each cell
separately. We tried other numbers of cells, and they give similar results.
In general, using a larger number of cells gives slightly better fits,
but we choose 32 cells as a compromise so as not to be in danger of
overfitting (i.e., having too few objects per cell for a 15-parameter fit). 
The three-dimensional kd-tree method gives a better result
than the method of cutting the color-magnitude plane into 10 regions
(see right panel, Figure~\ref{photoz_comp}).

Figure~\ref{photoz_comp}
compares the quality of the photometric redshifts obtained using 
the three different cutting methods described above. 
The panels from left to right are the photometric redshift vs.\
spectroscopic redshift diagrams obtained using:
(1) brute force single fit for all data, (2) cutting into 10
regions in the color-magnitude plane, and (3) the kd-tree method with 32 cells.
Both the high-$z$ and low-$z$ ends are improved with much reduced
systematics as we go from method (1) to method (3). The scatter in the
differences between photometric and spectroscopic redshifts 
is reduced as more advanced cutting algorithms are used. Hence, we
choose the three-dimensional kd-tree algorithm with 32 cells for
our final photometric redshift catalog. 
 
\section{Photometric Redshift Error}\label{photoz-err}
By examining the distribution of the differences between spectroscopic
redshifts and photometric redshifts for the training set, we can
estimate the uncertainty in the redshift fits. However, this does not
provide a measurement of the photometric redshift error for individual
galaxies in the catalog.  Knowing the confidence limits on the
photometric redshift measurements for individual objects is very
important for subsequent science analyses. Without the confidence
limits for each object, the analyses based on the photometric redshift
may suffer from catastrophic errors, which happen when the photometric
redshift estimates are unknowingly very different from the true
spectroscopic redshifts.  By considering the photometric redshift
error for each object and taking it into account in estimating the
errors in a subsequent science analysis, one will obtain more
realistic confidence limits in the analysis results. In the following
subsections, we describe the photometric redshift error determined by
comparing the photometric redshifts to the spectroscopic redshifts in
our training set (empirical error), and we also describe the method we
use to estimate the photometric redshift error for individual objects
(computed error).

\subsection{Empirical Error}
\begin{figure}[t]
\includegraphics[angle=270,width=0.9\textwidth]{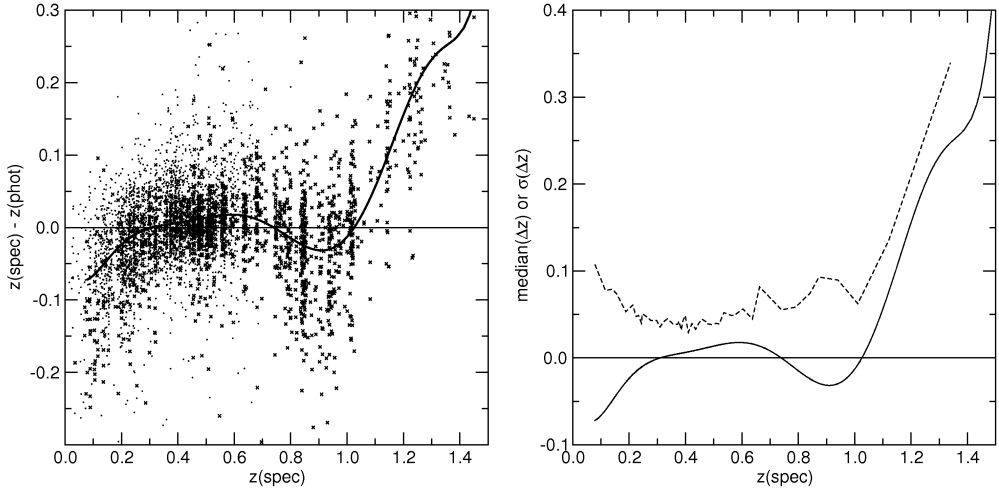}
\caption{$\Delta{z}$ vs. spectroscopic redshift diagram
(left), and median($\Delta{z}$) or $\sigma(\Delta{z})$ vs. 
spectroscopic redshift diagram (right) 
before noise is added. The fitting uses the kd-tree 32-cell cutting of the
training set. The dots indicate the RCS/CNOC2 data, the crosses 
indicate the GOODS/HDF-N data, and the solid curve in the left panel is 
the median value of $\Delta{z}$. The same solid curve is also shown in the
right panel for clearer presentation. The dashed curve in the right panel
is the $\sigma(\Delta{z})$ which indicates the 68th 
percentile difference between the photometric redshift and the 
spectroscopic redshift, in bins of 100 objects each
along the spectroscopic redshift axis. The $\sigma(\Delta{z})$ for objects 
at our median redshift is less than 0.05. The error becomes larger 
($\sim 0.09$) for objects at $z \sim 0.9$. 
The extremely large error for objects at $z > 1.1$ is due to the 
large systematic deviations at those high redshifts.
\label{zerr} }
\end{figure}

\begin{figure}[t]
\includegraphics[angle=0,width=0.9\textwidth]{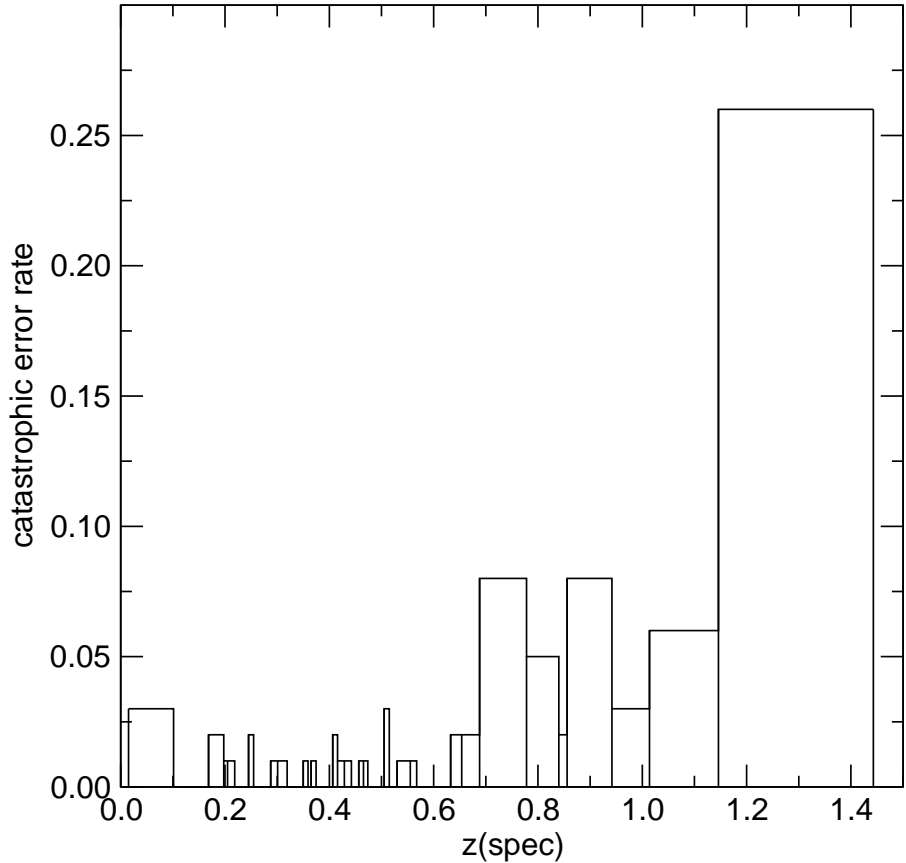}
\caption{Histogram of catastrophic error rate vs. spectroscopic redshift.
We define the catastrophic error rate as the ratio of the number of objects 
with $\mid\Delta{z}\mid > 0.5$
to the total number of objects for each redshift bin. The result is
calculated using the RCS/CNOC2 and noise added GOODS/HDF-N data. 
Note that the histogram has variable redshift bin widths because
we choose widths that always include exactly 100 objects per bin.
The catastrophic error rate is below
0.03 for $z < 0.7$. It becomes around 0.05 for $0.7 < z < 1.2$ due to larger
$\sigma(\Delta{z})$. 
For redshifts higher than 1.2, the catastrophic error rate is
0.25, which is primarily due to the larger systematic error.
\label{catastrophic_err}}
\end{figure}

Figure~\ref{zerr} presents the accuracy of the photometric
redshifts by comparing the photometric redshift and the
spectroscopic redshift for the objects in the training set. 
The $\sigma(\Delta{z})$ is the 68th percentile difference between the
photometric redshift and the spectroscopic redshift in bins of 100 objects 
each along the spectroscopic redshift axis. This provides a general estimate of
how statistically accurate our photometric redshift measurements are.
For objects at $0.20 < z < 0.65$, the rms scatter $\sigma(\Delta{z})$
is less than 0.05. The overall rms scatter is 0.068.
However, most objects at $z > 0.7$ in the training set are from
the GOODS/HDF-N sample, which have a much deeper limiting magnitude.
The photometric redshift error for an object
in the RCS at high-$z$ is thus under-estimated by the $\sigma(\Delta{z})$ 
of the training set. By adding additional Gaussian
noise to the GOODS/HDF-N data, we simulate the photometric
redshift error as if the GOODS/HDF-N data have the same depth and seeing
conditions as the RCS data. The results are shown in Figure~\ref{zerr1}. 
For objects at $0.2 < z < 0.5$, the rms scatter remains roughly the same, 
but it becomes $\sim 0.2$ for an object at $z \sim 0.8$. The 
extremely large rms for objects at $z > 1.1$ is caused by the large 
systematic deviation of the photometric redshifts from the spectroscopic
redshifts. The overall average rms scatter is 0.11. This result shows
the real error levels in the photometric redshift catalog.
We also calculate the catastrophic error rate for different redshift ranges
in bins of 100 objects each
and show the result in Figure~\ref{catastrophic_err}. We define the
catastrophic error rate as the ratio of the number of objects with 
$\mid\Delta{z}\mid > 0.5$
to the total number of objects for each redshift bin. The result is
calculated using the RCS/CNOC2 and noise added GOODS/HDF-N data. 
Note that the histogram has variable redshift bin widths because
we choose widths that always include exactly 100 objects per bin.
The catastrophic error rate is below
0.03 for $z < 0.7$. It becomes around 0.05 for $0.7 < z < 1.2$ due to larger
$\sigma(\Delta{z})$. 
For redshifts higher than 1.2, the catastrophic error rate is
0.25, which is primarily due to the larger photometric redshift 
systematic error. 

\subsection{Computed Error}
Basically the photometric redshift uncertainty for an individual object comes 
from two sources. One source is the error in the photometric data themselves, 
and the other source is the uncertainty in the empirical fitting 
parameters in the quadratic formula. As described below, we will 
use the combination of a Monte-Carlo method and a bootstrap method 
to estimate the effect of both these error sources and thereby compute the 
total photometric redshift error for each object. We refer to this error 
as the ``estimated'' or ``computed'' error.

The measurement error of the photometry is estimated from two sources.
One is from the uncertainty of the sky measurement; the other is from the
internal error of the magnitude measurement. The photometry errors are given
by PPP \citep{yee1991}. The Monte-Carlo method provides a way to estimate 
how the photometric errors statistically propagate into the error on the 
photometric redshift. By assuming that the distribution of the photometric 
uncertainties is Gaussian, we generate 500 photometric measurements,
using a normal distribution 
with a width equal to the photometric error and centered on the original 
magnitude in each filter. For the estimates of the uncertainties in the
photometric redshift fitting parameters (Equation~\ref{photozeqn}), the 
training set is bootstrap-resampled 500 times to generate 500 bootstrapped 
training sets and corresponding solutions. By combining 500 simulated 
photometric data sets with 500 bootstrapped 
training sets, 250,000 photometric redshift estimates are produced for 
each galaxy.
The median value of these 250,000 redshifts is chosen to be the photometric 
redshift of the galaxy, and the 68\% width ($\sim \nsig{1}$ if the photometric 
redshift error distribution is Gaussian) is the estimated photometric 
redshift error. Note that the photometric redshift error distribution is not
necessarily Gaussian. Our computed error estimate is intended to cover
the 68\% error range, regardless of the shape of the photometric redshift
error distribution.

\begin{figure}[t]
\includegraphics[angle=0,width=0.9\textwidth]{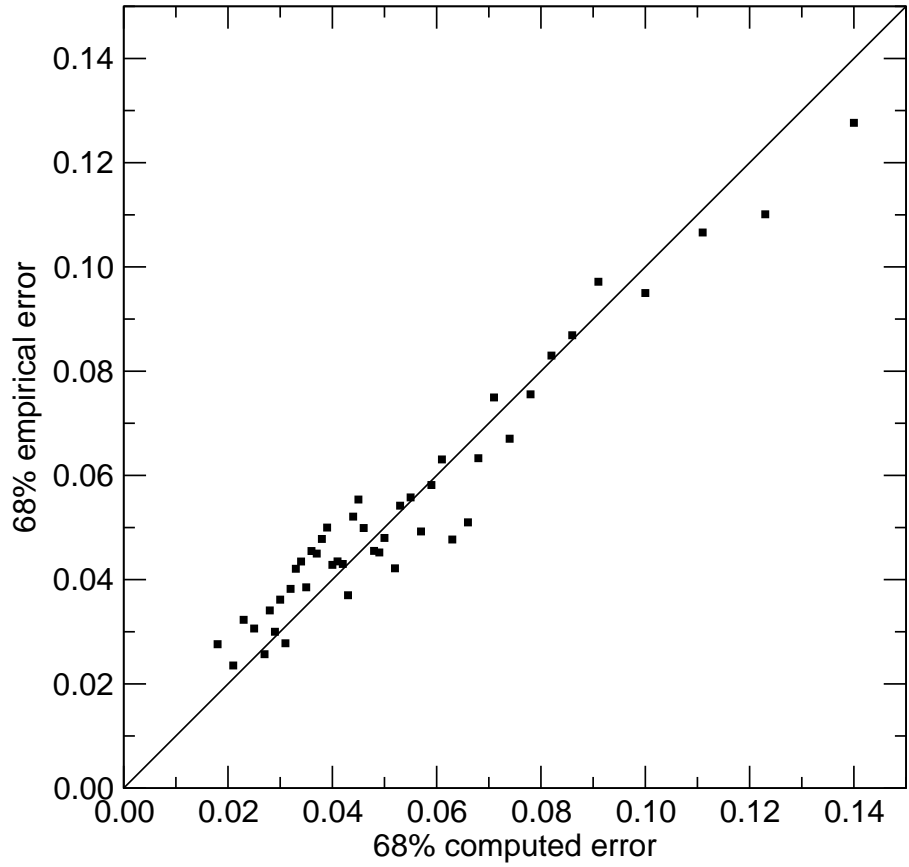}
\caption{The empirical error vs. the computed error for RCS and GOODS/HDF-N
data in our training set. The empirical error is the 
68th percentile difference between the photometric redshift and the 
spectroscopic redshift, in bins of 100 objects each along the 68\% estimated 
error axis. The computed error of the data point is the median value 
of the computed errors in the same bin of 100 objects. The computed errors
agree with the empirical errors very well.
\label{err_comp}}
\end{figure}

To test the quality of our photometric redshift error estimates,
we compare the computed errors with the empirical errors for the galaxies
in the training set. Figure~\ref{err_comp} shows the comparison of the 
empirical error and the computed error for the RCS and GOODS/HDF-N data in our 
training set. The empirical error plotted is the 68th percentile difference 
between the
photometric redshift and the spectroscopic redshift, in bins of 100 objects 
each along the 68\% computed error axis. The computed error plotted 
is the median value of the computed errors of objects in the same bin of 
100 objects. The computed errors agree very well with the empirical errors, 
with $\mid$empirical error - computed error$\mid \sim 0.01$. 
This near exact agreement
of the empirical and computed error is fortuitous. When different methods
of separating the training set are used, in general there is an offset and/or
scaling of the computed error relative to the empirical error. However, in
general, regardless of the exact fitting algorithm used, the computed errors
are always well-behaved linear functions of the empirical error, showing that
our method of estimating individual errors is robust.

\section{Result}\label{result}
\begin{figure}[t]
\includegraphics[angle=270,width=0.9\textwidth]{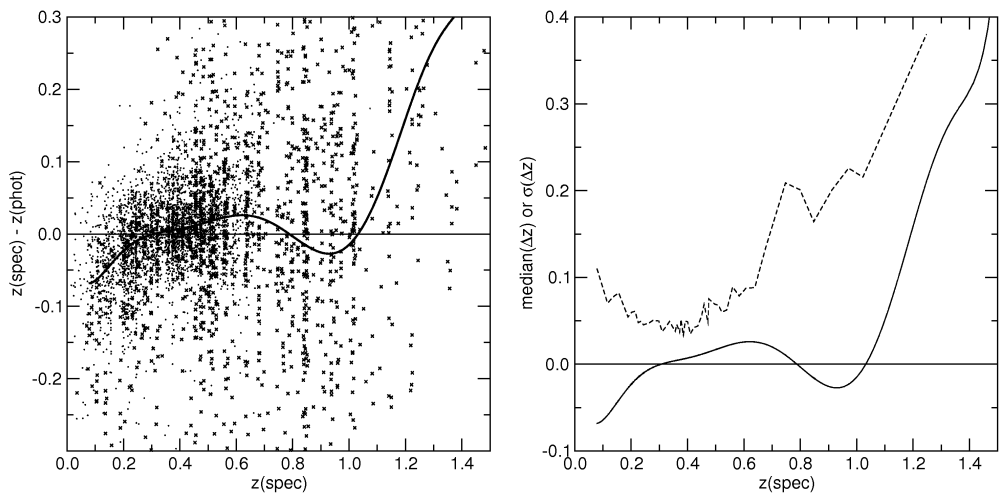}
\caption{This figure is the same as Figure~\ref{zerr} but noise has been added
to the GOODS/HDF-N sample (see text). The $\sigma(\Delta{z})$ for objects 
at our median redshift is similar compared to that in Figure~\ref{zerr},
but the error increases to around 0.2 for objects at $z > 0.8$, which 
is much larger than the value from Figure~\ref{zerr}. \label{zerr1}}
\end{figure}

To produce the final photometric redshift catalog, we use the 32-cell
kd-tree cutting method.
From Figure~\ref{zerr1}, for objects at $0.2 < z < 0.5$, the rms
scatter of $\Delta{z}$ is less than 0.06. For objects at $z > 0.8$, 
the rms scatter of $\Delta{z}$ is $\sim 0.2$. 
The overall average rms scatter is 0.11.
It can be seen that objects with lower spectroscopic redshifts are 
over-estimated while objects with higher spectroscopic redshifts are 
under-estimated (Figure~\ref{zerr1}). The bluest filter for 
the RCS is $B$({4500\AA}) and it is not blue enough for
the {4000\AA} break at $z < 0.2$. This creates a poor redshift estimate
for low-z ($z < 0.2$) objects, and they tend to have higher photometric 
redshifts than real redshifts because their SEDs and the SEDs of objects 
at $0.2 < z < 0.3$ are very similar. For those objects with redshift 
higher than 1.2, the {4000\AA} break is moving out of the reddest filter 
($z'$). A behavior opposite to the low-z objects is seen for the high-$z$ 
objects, i.e., they tend to have lower photometric redshifts than real 
redshifts.

\begin{figure}[t]
\includegraphics[angle=0,width=0.9\textwidth]{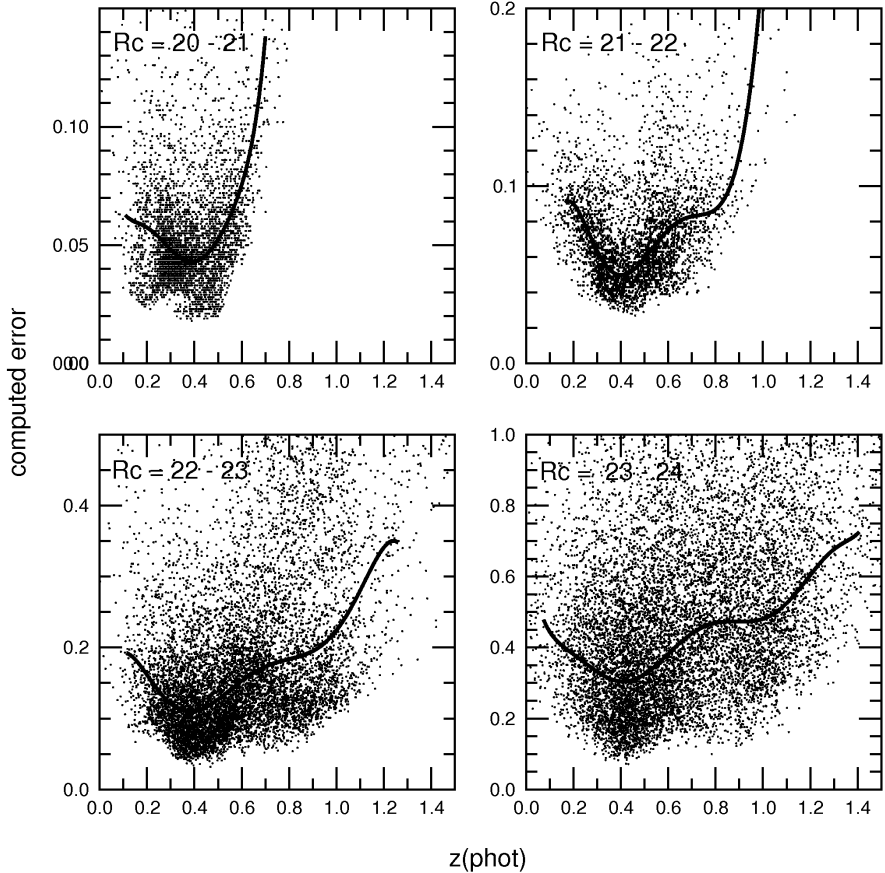}
\caption{Computed error vs. photometric redshift in different $R_c$ bins.
We plot a dot for every 15 and 20 objects for the $R_c = 20-21$ and 
$R_c = 21-22$ bins, respectively.
For the panels with $R_c = 22-23$ and $R_c = 23-24$ bins, we plot a dot for 
every 30 and 50 objects, respectively. The curve in each panel is the median 
value for all data points in each $R_c$ bin, not just for the data shown in the
figure. Note that we use different scales for the computed error axes in each
panel. The error grows with fainter $R_c$ and higher redshift. For the
objects at $z < 0.25$, the errors also become larger because we do
not have a sufficiently blue filter for low redshift galaxies.
Generally speaking, the photometric redshift error increases with magnitude 
due to larger photometric uncertainty.
\label{esterr_photoz}}
\end{figure}

We present the computed error vs. photometric redshift in different 
$R_c$ bins in Figure~\ref{esterr_photoz}. For the panels with 
$R_c = 20-21$, we plot a dot for every 15 objects. For the panel with
$R_c = 21-22$, we plot a dot for every 20 objects. For the panel with 
$R_c = 22-23$, we plot a dot for every 30 objects, and for the panel with 
$R_C = 23-24$, every 50 objects. The curve in each panel 
is the median value for all the data points in each $R_c$ bin, not just for 
the data shown in the figure. Note the different scales for the
computed error axes in each panel.
In general, for galaxies with $R_c < 22$ with photometric redshift
less than 0.8, the errors are below 0.1. For objects with $R_c = 22-23$,
the errors become larger but there are still a significant number of galaxies
with errors less than 0.1. For objects with $R_c = 23-24$, the photometric
redshift errors are greater than 0.3, which is due to larger photometric errors
and also to a higher fraction of late-type galaxies. Generally speaking, the
photometric redshift error increases with magnitude due to larger photometric
uncertainty.

\begin{figure}[t]
\includegraphics[angle=0,width=0.9\textwidth]{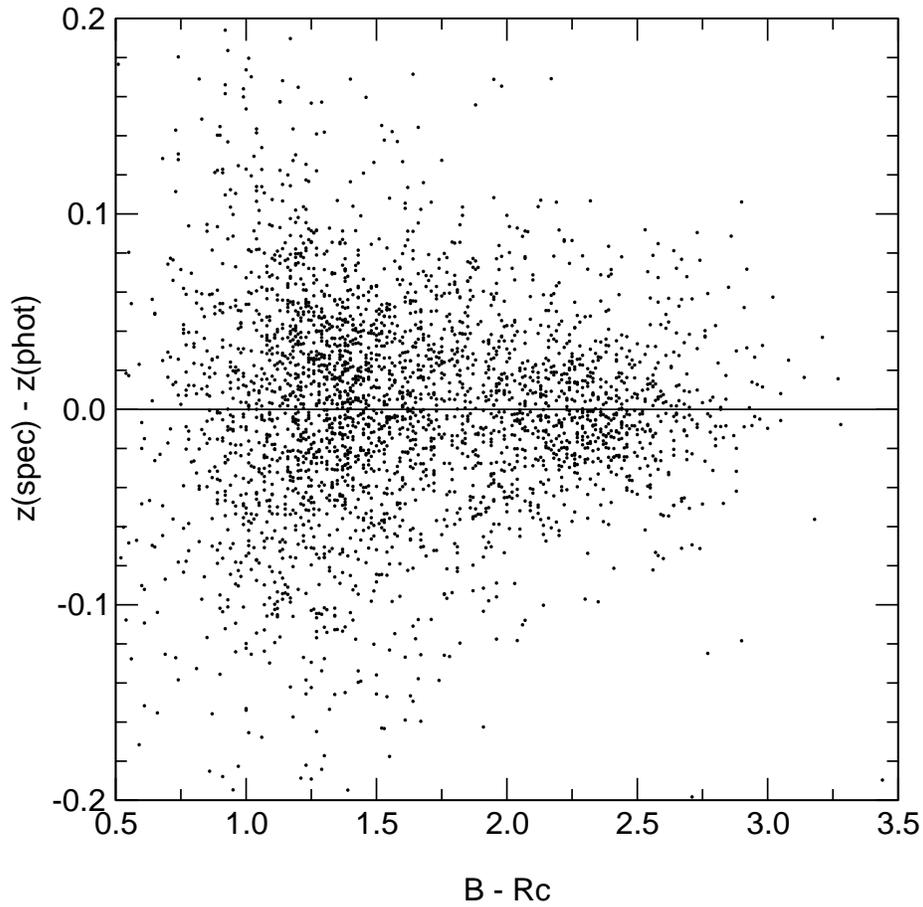}
\caption{$\Delta{z}$ vs. color ($B - R_c$) diagram for RCS data in
our training set.
$\sigma(\Delta{z})$ is 0.03 for redder objects ($B - R_c > 2$).
The width of the error distribution becomes 0.043 
for objects with $1.5 < B - R_c < 2$.
For objects with $B - R_c < 1.5$, this becomes larger than 0.055.
\label{color-err}}
\end{figure}

Although the galaxies are fit separately in several color and magnitude
bins, the rms scatter for bluer galaxies is still much larger than the
scatter for redder galaxies (shown in Figure~\ref{color-err}). This
is because early-type (redder) galaxies have some significant
spectral features (e.g., the {4000\AA} break) and have similar SEDs,
while late-type (bluer) galaxies tend to have a featureless and
relatively flat continuum. Ongoing star formation,
gas absorption/emission, and dust extinction will also
complicate the SEDs of late-type galaxies, and different galaxies
have different combinations of these effects. Because different
late-type galaxies have weaker features and larger variation
in their SEDs than early-type galaxies, the errors in the
photometric redshift estimates will be larger for late-type
galaxies.

We compare the photometric redshift technique we use to other techniques
described in \citet{csabai2003}. They use nine different methods including
template fits and empirical fits to estimate photometric redshifts
for the SDSS Early Data Release. From their study, in general, the results 
using empirical fits have smaller $\sigma(\Delta{z})$ than the 
ones using 
template fits, and the best algorithm with the smallest $\sigma(\Delta{z})$
among the empirical fits is also the kd-tree method. 
However, the kd-tree method 
they use is a two-dimensional tree to cut their training set in a color-color 
plane, unlike what we use, which is a three-dimensional tree in a color-color-magnitude
space. The addition of magnitudes provides a rough estimate of 
redshift which makes the
separation of the training set more precise than using just colors.
Combining deeper photometry, a larger training set over a wide redshift range,
and a better cutting method for the training set, 
the limiting magnitude of our photometric redshift result is $1.5 - 2.0$ 
magnitudes deeper than the SDSS result,
even with only four wide-band
filters (compared to five filters for SDSS) and a much higher redshift limit.
However, qualitatively, the agreement
between spectroscopic and photometric redshift is similar between the SDSS and
the RCS; e.g., the very low redshift galaxies tend to have over-estimated
photometric redshifts.

\begin{figure}[t]
\includegraphics[angle=0,width=0.9\textwidth]{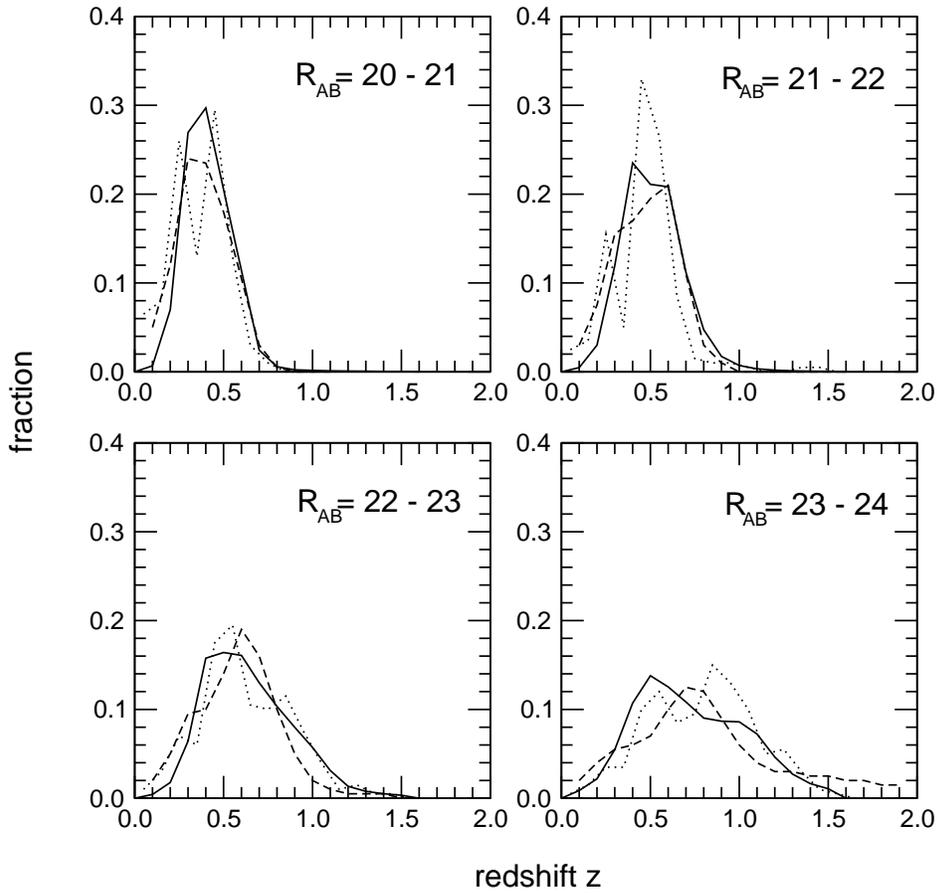}
\caption{The redshift distribution comparison diagrams. The solid
lines, dashed lines, and dotted lines indicate RCS photometric redshift, 
CFDF photometric redshift,
and GOODS/HDF-N spectroscopic redshift data, respectively. 
Overall the agreement between
the different samples is very good. \label{zdist}}
\end{figure}

To test our photometric redshift results, we compare the redshift
distributions of our data against the photometric redshift
distributions from the Canada-France Deep Fields
(CFDF)\citep{brodwin2003}, and the spectroscopic redshift
distributions for the GOODS/HDF-N field
\citep{wirth2004,cowie2004}. Figure~\ref{zdist} presents the
comparison results. The solid lines indicate the RCS data.
The dashed lines represent the CFDF data, and the dotted lines
indicate the GOODs data. The agreement between the different
samples is in general very good. The scatter for the GOODS data in the
$R_{AB} = 20-22$ panels is due to the small area of the
GOODS/HDF-N field ( $<$ 160 arcmin$^{2}$). The agreement for the three 
data sets is good in the $R_{AB} = 22-23$ bin. In the $R_{AB} = 23-24$ 
bin, the RCS data is somewhat broader presumably due to the larger redshift 
uncertainty, and it also has a slightly lower average. 
These comparisons provide a high level of confidence that the RCS photometric
redshift measurements, despite the relatively shallow photometry, are
statistically robust and reliable to as faint as $R_c \simeq 23$, whereas for
$23 < R_c < 24$, the photometric redshift catalog should be used with
caution.

\begin{figure}[t]
\includegraphics[angle=0,width=0.9\textwidth]{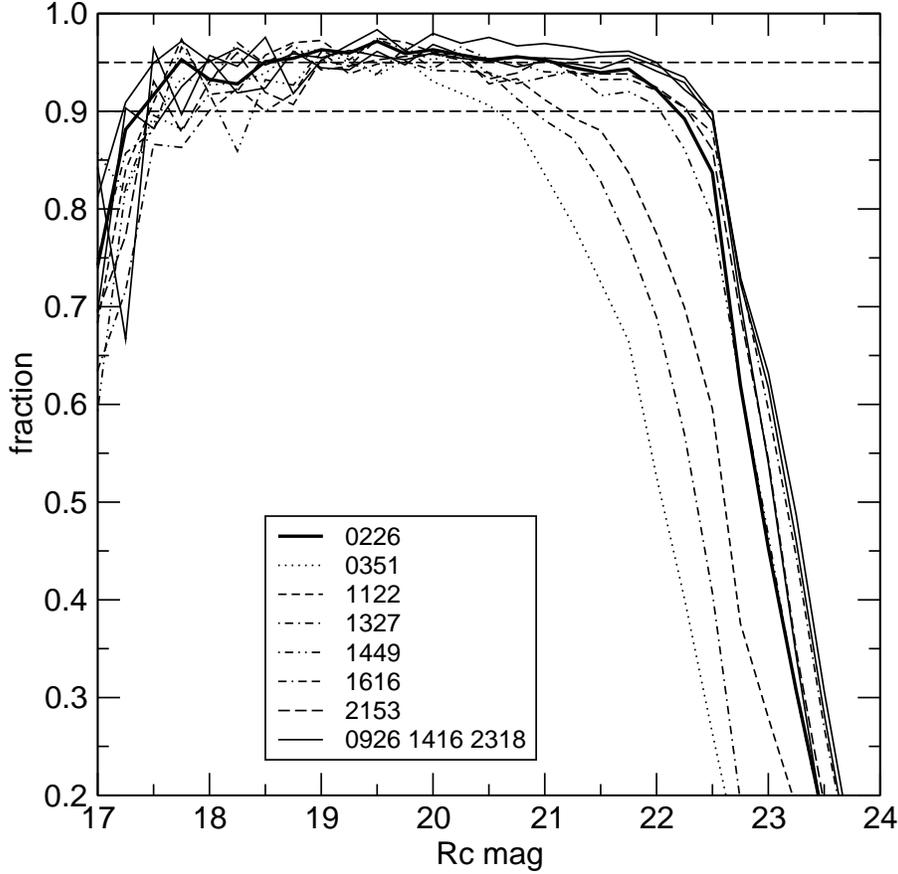}
\caption{
``Good'' photometric redshift fraction vs. $R_c$, defined as
$\sigma_z/(1+z) < 0.15$, where $\sigma_z$ is the computed photometric redshift
error. Each line indicates one patch. Patches 0351, 1122, and 1327 have
much lower fractions at the faint end due to the poorer $V$ and $B$ data
quality (see text and Figure~\ref{vb_lim}). 
For patches 0226, 1449, 1616, and 2153, the fractions are higher than
90\% for $17.5 < R_c < 22.0$. For the remaining three patches, the fractions
are higher than 90\% for $17.5 < R_c < 22.5$. The completeness levels
increase to $R_c \sim 23$ mag when the criterion 
$\sigma_z/(1+z) < 0.35$ is used.
\label{photoz-ratio}}
\end{figure}

All galaxies with $R_c < 24.0$ are included in the photometric
redshift catalog. The total number of galaxies is more than
1,300,000. About 168,000 objects (13\% of the total) have photometric
redshifts at $z < 0.0$ or $z > 1.5$, which are beyond the redshift
range of the training set. Negative redshifts are unlikely to be physical, 
and because the extrapolation of the polynomial fit is very unstable,
photometric redshifts $> 1.5$ cannot be trusted. For these objects,
we use a value of ``99'' to indicate this type of problematic photometric 
redshift in the catalog. Figure~\ref{photoz-ratio} presents the
fraction of good photometric redshifts vs. $R_c$ magnitude. 
The definition of ``good photometric redshift'' is $\sigma_z/(1+z) < 0.15$,
where $\sigma_z$ is the computed photometric redshift error. Patches 
0351, 1122, 
and 1327 have much poorer success rates at the faint end compared to the
other patches due to poor $V$ and $B$ data quality 
(the limiting magnitudes for 
$V$ and $B$ are about 24 and 23.5, respectively, compared to the average of 
24.5 and 25.0, see Figure~\ref{vb_lim}). 
For galaxies with $17.5 < R_c < 22.0$ in patches 
0226, 1449, 
1616, and 2153, the fractions of good redshifts are higher than 90\%. For the 
remaining four patches, the fractions are roughly higher than 90\% for 
galaxies with $17.5 < R_c < 22.5$. For brighter ($R_c < 17.5$) 
galaxies, the lower fractions are probably due to a lack of bright galaxies in 
our training set, improper star-galaxy separation due to saturated stars, 
saturation of bright galaxy images in one filter or more, 
and the lack of a sufficiently 
blue filter for galaxies at low redshift. For fainter galaxies, the fractions 
are also going down due to larger magnitude errors, especially for
redder galaxies because they are even fainter in $V$ and $B$
magnitudes. According to this figure, an overall reliable completeness 
magnitude limit (90\% to 98\%) for the full catalog is $17.5 < R_c < 20.5$,
while if the patches 0351, 1122, and 1327 are removed, the statistical 
completeness limit extends to $R_c \sim 22$. If we relax the photometric 
redshift uncertainty limit to $\sigma_z/(1+z) < 0.35$, the completeness limit
extends to $R_c = 23$.

\begin{figure}[t]
\includegraphics[angle=270,width=0.9\textwidth]{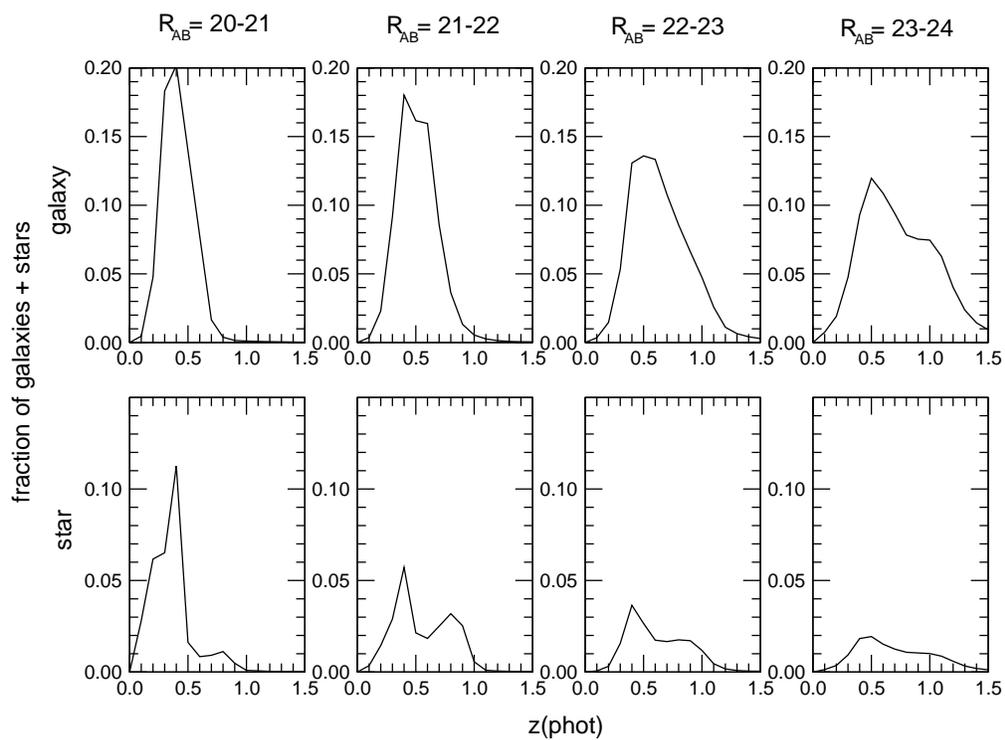}
\caption{Redshift distributions for PPP classified galaxies and stars
in different $R_{AB}$ bins. The upper four panels represent the galaxy
fraction among all objects (galaxies + stars). The lower four panels show
the ``stars'' fraction among all objects. Only for the $R_{AB} = 23-24$ bin is 
the redshift distributions very similar between the upper and lower panels. 
It implies that many ``stars'' in this magnitude range are actually galaxies.
\label{zdist.gal_star}}
\end{figure}

Because PPP star-galaxy classification begins to become less robust at
$R_c \geq 22$ due to the low signal-to-noise ratio for faint
objects \citep{yee1991}, some faint galaxies are classified as stars.
To test the effect and extent of misclassification, a ``fake'' photometric 
redshift catalog for point sources in each patch is also generated. 
If we force fit photometric redshifts to real stars, it would produce 
results that systematically do not have the same redshift distribution as 
the galaxies. Figure~\ref{zdist.gal_star} demonstrates the photometric
redshift distributions of PPP classified galaxies and stars in different
$R_{AB}$ magnitude bins. The upper panels show the galaxy fraction of
all objects (galaxies + stars), and the lower panels show the stellar fraction.
For the $R_{AB} = 20-21$, $21-22$, and $22-23$ bins,
the redshift distributions for PPP classified galaxies and stars are
not similar. The ``fake'' photometric redshifts of stars tend to be
separated into two bins at $z \sim 0.4$ for the bluer disk stars, and
$z \sim 0.8$ for the redder faint halo stars. The peaks for the three
brighter bins do not change with magnitude, unlike those for the galaxies.
But for the $R_{AB} = 23-24$ bin, the shape of the redshift distributions in 
the upper and lower panels are very similar, which implies that a significant 
fraction of the ``stars'' are actually galaxies. Of course, we cannot
rule out that high-surface brightness, concentrated galaxies
(more likely to be misclassified as stars) may have a different redshift
distribution compared to low-surface brightness, less concentrated galaxies.

Table~\ref{catalog} is an example showing the information available for
each object in the photometric redshift catalog; the full photometric redshift
catalog (Table~4) is available in the electronic version of the paper. Listed
in the sample catalog are the following rows:

$Row (1).$\textemdash The RCS object number. 
The first two digits indicate the patch number, numbered 01 to 10 in the order
of the patches in Table~\ref{patch}.
The second two digits indicate the pointing number within a patch.
The third two digits indicate the CCD chip number within the pointing. 
The remaining five digits represent the object number within each CCD chip.

$Rows (2)-(3).$\textemdash The R.A. in hours and Dec. in degrees (J2000).

$Rows (4)-(15).$\textemdash The total magnitudes, magnitude errors, and 
color errors for $z'$, $R_c$, $V$, and $B$. The magnitude errors are derived 
from the photon noise in the optimal aperture 
(on total magnitude, see \citealt{yee1991,GY2004}) for each object. The color 
errors are the sum in quadrature of the photon errors for each filter 
in the color aperture. 

$Row (16).$\textemdash The photometric redshift.

$Row (17).$\textemdash The computed error on the photometric redshift.

\clearpage

\begin{deluxetable}{cccccc}
\tabletypesize{\small} \tablecolumns{26} \tablewidth{0pt}
\tablecaption{Catalog Sample\label{catalog}} \tablehead{
\colhead{Object ID} & \colhead{01110800110} & \colhead{01110800128} &
\colhead{01110800147} & \colhead{01110800158} & \colhead{01110800174}}
\startdata 
RA & 2.435250  & 2.428926 &
2.432872 & 2.429133 & 2.432602 \\
DEC & 0.90261 & 0.90425 &
0.90534 & 0.90611 & 0.90704 \\
$z'$ & 20.81 & 18.70 &
20.04 & 19.27 & 19.43 \\
+/- m, +/- c of $z'$ & 0.05 0.04 & 0.01 0.01 &
0.03 0.02 & 0.02 0.01 & 0.01 0.01 \\
$R_c$ & 21.00 & 19.36 &
20.58 & 19.98 & 20.04 \\
+/- m, +/- c of $R_c$ & 0.02 0.01 & 0.01 0.00 &
0.01 0.01 & 0.01 0.01 & 0.01 0.01 \\
$V$ & 21.45 & 20.67 &
21.66 & 21.28 & 21.30 \\
+/- m, +/- c of $V$ & 0.04 0.02 & 0.01 0.01 &
0.03 0.03 & 0.03 0.02 & 0.02 0.02 \\
$B$ & 22.23 & 22.06 &
22.79 & 22.58 & 22.67 \\
+/- m, +/- c of $B$ & 0.03 0.03 & 0.03 0.03 &
0.05 0.05 & 0.05 0.04 & 0.04 0.04 \\
Photometric Redshift & 0.263 & 0.436 &
0.393 & 0.442 & 0.395 \\
Redshift error & 0.072 & 0.031 &
0.071 & 0.031 & 0.018 \\
\enddata
\end{deluxetable}

\clearpage

\section{Conclusion}\label{conclusion}
We present a photometric redshift catalog from the RCS data, constructed by
applying an empirical polynomial fitting technique using a training 
set of 4924 objects with redshifts from the CNOC2 and GOODS/HDF-N samples.
A 32-cell kd-tree algorithm is used to divide up our sample to improve
the accuracy of the photometric redshift estimates.
Our catalog includes 1.3 million galaxies in 33.6 deg$^{2}$ 
of sky (distributed over 10 patches) with redshifts less than 1.5.
The rms photometric redshift scatter is $\sigma(\Delta{z}) < 0.06$ 
within the redshift range $0.2 < z < 0.5$, and 
$\sigma(\Delta{z}) < 0.11$ for galaxies with $0.0 < z < 1.5$.  
The computed photometric redshift errors 
for individual galaxies are also provided. The magnitude limit for 
completeness of the catalog is $17.5 < R_c < 22$ (with 
$\sigma_z/(1+z) < 0.15$) if the three shallow
patches (0351, 1122, and 1327) are excluded. 
The limit extends to $R_c \sim 23$ for $\sigma_z/(1+z) < 0.35$.
We also compare the redshift distribution from our catalog to 
the CFDF and the GOODS/HDF-N data, and the agreement between the 
different samples is in general very good.

We are carrying out a number of scientific studies using the photometric
redshift catalog. One important example is to provide a large sample of
galaxies over a large look-back time to measure the luminosity function and its
evolution for field galaxies (Lin et al.~2005, in preparation). Moreover,
the luminosity function for galaxies in clusters discovered by the RCS 
can be determined more accurately using photometric redshifts than by using 
two filters ($R_c$ and $z'$). Furthermore, we plan to develop cluster
finding algorithms using the photometric redshift catalog, creating cluster
catalogs for comparison with the result from the RCS technique 
\citep{GY2000,GY2004}.
This will allow us to verify the completeness of the two-filter cluster
red-sequence finding algorithm.
Another topic is the search for close galaxy pairs, both in the field and 
in galaxy clusters, to study the evolution of the merging rate 
(Hsieh et al.~2005, in preparation). Such a study will provide data
to test the hierarchical model of galaxy evolution. We are also using the
catalog to find sub-structures (groups) in clusters, and to study the cluster
population as a function of radius. Finally, the photometric redshift 
catalog provides accurate redshift distributions of lens galaxies and 
source galaxies, which are both very important for weak lensing 
analysis. With photometric redshift information, many weak lensing analyses,
from galaxy-galaxy lensing to cosmic shear, can be improved. For example,
knowing the redshifts of the lenses allows one to derive the mass-to-light
ratio of galaxies as a function of galaxy luminosity
(Hoekstra et al.~2005, in preparation), testing the scaling relations of 
bayonic and dark matter. 

\acknowledgments
We wish to thank the staff at CFHT for the observations and reduction of
part of the data under the queue system. The research of H. Y. is supported
by grants from the National Science and Engineering Research Council of Canada
and the University of Toronto.

\end{document}